\documentclass[twocolumn,preprintnumbers,amsmath,amssymb]{revtex4}

\usepackage{graphicx}
\usepackage{subfigure}
\usepackage{dcolumn}
\usepackage{bm}
\begin{document}
\title{\bf{The Effect of Symmetry Lowering on the Dielectric Response
of BaZrO$_{3}$} \\[11pt] } \author{Joseph W. Bennett, Ilya Grinberg,
Andrew M. Rappe} \affiliation{ The Makineni Theoretical Laboratories,
Department of Chemistry\\ University of \ Pennsylvania, Philadelphia,
PA 19104-6323 } \date{\today}

\begin{abstract}
We use first-principles density functional theory calculations to
investigate the dielectric response of BaZrO$_3$ perovskite.  A
previous study [Arkbarzadeh {\em et al.} Phys. Rev. B {\bf 72}, 205104
(2005)] reported a disagreement between experimental and theoretical
low temperature dielectric constant $\epsilon$ for the high symmetry
BaZrO$_3$ structure.  We show that a fully relaxed 40-atom BaZrO$_3$
structure exhibits O$_6$ octahedral tilting, and $\epsilon$ that
agrees with experiment.  The change in $\epsilon$ from high-symmetry
to low-symmetry structure is due to increased phonon frequencies as
well as decreased mode effective charges.
\end{abstract}

\maketitle
\section{\label{sec:level1} Introduction}
 Dielectric materials are important for wireless communications
 technology.  These devices require a high dielectric constant,
 $\epsilon$, and low dielectric loss~\cite{Fortin96p4237}. Barium
 zirconate, BaZrO$_{3}$ (BZ), is one of the constituent materials in
 the electroceramic capacitors used in wireless communications.  BZ is
 a classic ABO$_3$ perovskite dielectric material that is both
 chemically and mechanically stable.  

According to the well-established tolerance factor argument, BZ should
 have a stable cubic structure.  Tolerance factor $t$ is given by
\begin{eqnarray}
 t = \frac {R_{\rm A-O}}  {R_{\rm B-O} \sqrt 2}
\end{eqnarray}            

\noindent where $R_{\rm A-O}$ is the sum of A and O ionic radii and
$R_{\rm B-O}$ is the sum of B and O ionic radii.  Tolerance factor
$t<1$ usually leads to the rotation and expansion of the B-O$_6$
octahedra. Such octahedral rotations often generate a low temperature
anti-ferroelectric (AFE) phase (e.\ g.\ PbZrO$_3$).  If $t>1$, the
B-O$_6$ octahedra are stretched from their preferred B-O bond lengths,
promoting B-cation distortions by creating room for the B-cations to
move off-center.  Therefore, simple perovskites with $t>1$ are usually
ferroelectric (FE).  For BZ, the sizes of the Ba and Zr ions ($R_{\rm
Ba}$=1.61~\AA\ and $R_{\rm Zr}$=0.72~\AA), in their O$_{12}$ and
O$_{6}$ cage ($R_{\rm O}$=1.35~\AA) exactly balance, leading to t~=~1,
so the cubic structure should have no driving force to deform to a
lower symmetry. It also has a stable dielectric constant over a wide
range of temperatures~\cite{Levin03p170}. 

The total dielectric constant is a sum of two contributions, related
to the electronic ($\epsilon_{\infty}$) and ionic polarizabilities
($\epsilon_{\mu}$). For most perovskites, $\epsilon_{\infty}$ is small
($\approx$ 5) and $\epsilon_{\mu}$ is the dominant contribution at low
frequency. Recent experimental studies of BZ at low temperature show
that $\epsilon$ is 47 at 0~K and some interesting quantum effects are
present, though not large in
magnitude~\cite{Akbarzadeh05p205104}. Reference 3 also presents a DFT
study which calculates $\epsilon$ of BZ to be 65 at 0~K,
overestimating experimental results. DFT inaccuracies have been known
for some time in describing ferroelectric systems~\cite{Cohen96p1393}.
It is however surprising that DFT is inaccurate for a system as simple
as paraelectric BZ.

Here, we show that at low temperatures, the ground state structure of
BZ is not accurately represented by a five atom cell. A larger
supercell with lower symmetry is energetically favored. The calculated
value of $\epsilon$ for this low symmetry structure is in agreement
with experimental results. Most importantly, the relationship between
the breaking of structural symmetry and the change in dielectric
response is elucidated.

\section{\label{sec:level1} Methodology}
In this study, two first principles codes are used. An in-house solid
state DFT code used in previous
studies~\cite{Grinberg03p130,Mason04p161401R} and the ABINIT
software package~\cite{Gonze02p478} are used to relax the ionic
positions and lattice constants.  Local density approximation (LDA) of
the exchange correlation functional and a $4\times4\times4$
Monkhorst-Pack sampling of the Brillouin zone~\cite{Monkhorst76p5188}
are used for all calculations. All atoms are represented by
norm-conserving optimized~\cite{Rappe90p1227} designed
nonlocal~\cite{Ramer99p12471} pseudopotentials. All pseudopotentials
generated using the OPIUM code~\cite{Opium}. The calculations are
performed with a plane wave cutoff of 50~Ry.

Once the structure is fully relaxed, response
function~\cite{Gonze97p10355,Ghosez98p6224} calculations are
performed with ABINIT to generate $D_{\alpha\beta}(i,j)$, the
mass weighted dynamical matrix.

\begin{eqnarray}
D_{\alpha\beta}(i,j)=\frac{\partial^2 E}{\sqrt{m_{i}m_{j}}\partial{\tau_{i\alpha}}\partial{\tau_{j\beta}}}.
\end{eqnarray}            

Once normalized, each eigenvalue is $\nu_{\mu}^{2}$, the frequency
squared of a normalized eigenvector $a_{\mu}$. Born effective charge
tensors, $Z_{i\alpha\beta}^{*}$ are also calculated for each
atom. Each mode $\mu$ has a mode effective charge,
$Z_{\mu\alpha}^{*}$, defined as
\begin{eqnarray}
Z_{\mu\alpha}^{*}=\sum_{i\beta}\frac{Z_{i\alpha\beta}^{*}(a_{\mu})_{i\beta}}{m_{i}^{1/2}},
\end{eqnarray}            

Contributions to the dielectric response arise only from the IR
active modes~\cite{Cockayne00p3735}. These modes have a non-zero
$Z_{\mu\alpha}^{*}$, and are used to calculate $\epsilon_{\mu}$, the
contribution to the dielectric constant from mode
$\mu$~\cite{Cockayne03p2375} as

\begin{eqnarray}
\epsilon_{\mu\alpha\beta}
=\frac{Z_{\mu\alpha}^{*}Z_{\mu\beta}^{*}}{4\pi^{2}\epsilon_{0}V\nu_{\mu}^{2}}.
\end{eqnarray}

where the total ionic contribution is

\begin{eqnarray}
\epsilon_{\mu}=\frac{1}{3}\sum_{\alpha}\epsilon_{\mu\alpha\alpha}
\end{eqnarray}

\section{\label{sec:level1} Results and Discussion}
\subsection{\label{sec:level2} Five-Atom BZ calculations}
\begin{figure}
\includegraphics[width=3.0in]{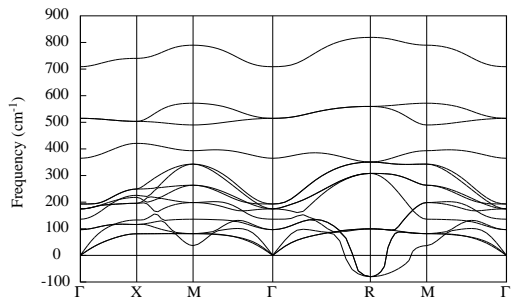}
\caption{{The calculated phonon band structure of the five-atom
high-symmetry BZ structure. Our results match well with the previously
published BZ phonon band structure in
reference~\cite{Akbarzadeh05p205104}. In particular, our unstable mode
at R is -79 cm$^{-1}$, in close agreement with their -63 cm$^{-1}$}}
\label{fig:BZBands}
\end{figure}

The ground state five atom BaZrO$_{3}$ is a cubic perovskite structure
where all ions occupy high symmetry positions. The relaxed DFT lattice
constant of this cell was $a$~=4.157 \AA. This structure was used to
calculate both a phonon band structure (Figure~\ref{fig:BZBands}) and
$\epsilon$. The high-symmetry BaZrO$_{3}$ structure contains three IR
active phonon modes in each direction.

\begin{table}
\begin{tabular}{ccc|c}
\hline
$\omega$&$Z^{*}_{\mu}$&$\epsilon_{\mu}$&Motion\\
\hline
96      &2.1     &40.8   &Ba,Zr-O\\
193     &2.8     &17.2   &Zr-O\\
513     &2.8     &2.4    &O$_{6}$\\
\hline
\end{tabular}
\caption{Forty-atom BZ phonon data for IR active modes in the
high-symmetry cubic structure. Frequencies are in cm$^{-1}$, mode
effective charges, $Z^{*}_{\mu}$, are in electrons. The contributions
to the $\epsilon$ are calculated in SI units. Also reported are the
motions which create the IR active mode. }
\label{table:BZ5PhonTable}
\end{table}

The first set of phonon frequencies, occurring at 96 cm$^{-1}$
contribute the most to $\epsilon$, as shown in
Table~\ref{table:BZ5PhonTable}. These three modes are mainly large Ba
displacements against its O$_{12}$ cage into which some Zr
displacement opposite its O$_{6}$ cage is mixed (Last mode).  The
ratio of Ba displacement to Zr displacement is about 9:1.  However,
due to the much larger $Z^{*}$ of Zr (6.3 for Zr versus 2.7 for Ba)
even such a small Zr off-center displacement makes a significant
contribution to the effective charge of the mode.  The IR active modes
are sometimes discussed in terms of purely A-O or B-O displacements,
but that simple cation motion is not justified when describing the
lowest frequency phonon modes of BZ.

The second set of phonon frequencies occurs at 193 cm$^{-1}$,
consisting of Zr motion (Slater mode). A small amount of Ba motion is
also present, but the ratio of Zr off-centering to Ba off-centering in
this mode is about 20:1.  Combined with the smaller $Z^{*}$ of Ba,
this means that this mode is well-characterized as pure B-O.  The
final set of phonons, at 513 cm$^{-1}$ are caused by asymmetric
O$_{6}$ cage motions. These contribute the least to $\epsilon$.

To obtain $\epsilon$, the contributions of $\epsilon_{\mu}$ from
Table~\ref{table:BZ5PhonTable} are added to $\epsilon_{\infty}$, which
was calculated as 4.94. The total zero temperature dielectric constant
for the high symmetrystructure is 65. This value is almost double
previous experimental investigations~\cite{Levin03p170}, but agrees
well with recent theoretical
investigations~\cite{Akbarzadeh05p205104}.

The disagreement between experimental and theoretical $\epsilon$
results suggests either an inaccuracy of LDA or an inaccurate
representation of ground-state BaZrO$_{3}$ structure by a five-atom
high-symmetry unit cell.  In Reference~\cite{Akbarzadeh05p205104}, the
authors present a phonon band structure of BZ, which displayed a soft
mode at R, but none at $\Gamma$. Our phonon band structure of five
atom BZ, presented in Figure~\ref{fig:BZBands}, confirms that there is
no ferroelectric softening at $\Gamma$, but a rather large softening
at R. The presence of an unstable mode indicates a driving force for
symmetry breaking. To explore the effect of this, a larger BZ supercell is
needed for accurate calculations.

\subsection{\label{sec:level2} Forty-Atom BZ calculations}
The soft mode at R observed in the phonon band structure was induced
in the high symmetry structure, and the supercell was then
relaxed. The relaxed low symmetry structure had lattice constants
$a$~=~8.306\AA, $b$~=~8.313\AA, and $c$~=~8.290\AA. Thus, the relaxed
volume is only 0.39$\%$ less than the high symmetry structure. None of
the cations shift from their high symmetry positions, consistent with
an O$_{6}$ tilt mode. Octahedral tilt angles range between 3.6 and
4.2$^{\circ}$. A subsequent phonon response calculation on the tilted
pseudo-cubic structure showed that no negative frequency modes were
present. Unlike our calculations, low temperature x-ray diffraction
and neutron scattering experiments are unable to detect this tilting
in BZ~\cite{Akbarzadeh05p205104}. Current experimental evidence is
unclear.  The relevant X-ray reflections for oxygen-based symmetry
breaking would be weak, due to the scattering factor of O. We propose
the tilted structure based on quantum mechanical energy minimization
and consequent dielectric response (see below) and we suggest further
experimental investigation to directly confirm the predicted
structure.

\begin{table}
\begin{tabular}{l|lll|lll|c}
\hline
$\omega$&$Z^{*}_{xx}$&$Z^{*}_{yy}$&$Z^{*}_{zz}$&$\epsilon_{xx}$&$\epsilon_{xx}$&$\epsilon_{zz}$&Motion\\
\hline 
106     &1.3    &1.3     &0.5     &12.1    &11.7    &1.9     &Ba-O\\
107     &0.5    &0.5     &0.2     &2.0     &1.9     &0.3     &Ba-O\\
114     &0.4    &0.4     &1.8     &1.0     &0.8     &21.8    &Ba-O\\
125     &1.2    &1.3     &0       &8.5     &9.0     &0       &Ba-O\\
\\
184     &0.2    &0.3     &0.1     &0.1     &0.2     &0.0     &Zr-O\\
192     &0.3    &0.7     &0       &0.3     &1.1     &0       &Zr-O\\
195     &1.3    &2.3     &1.0     &3.6     &12.2    &2.2     &Zr-O\\
196     &0.3    &1.0     &2.7     &0.2     &2.0     &16.3    &Zr-O\\
197     &2.5    &1.3     &0.2     &14.1    &3.5     &0.1     &Zr-O\\
211     &0.6    &0.6     &0       &0.7     &0.6     &0       &Zr-O\\
\\
307     &0.2    &0.1     &0.6     &0.1     &0.0     &0.3     &Ba,Zr-O\\
308     &0.4    &0.4     &0.2     &0.2     &0.1     &0.1     &Ba,Zr-O\\
\\
501     &1.5    &1.9     &1.2     &0.8     &1.2     &0.5     &O$_{6}$\\
502     &0.8    &0.9     &2.4     &0.2     &0.3     &1.9     &O$_{6}$\\
505     &2.1    &1.7     &0       &1.4     &0.9     &0       &O$_{6}$\\
\hline
\end{tabular}
\caption{Forty-atom low-symmetry BZ phonon data for IR active
modes. Frequencies are in cm$^{-1}$, mode effective charges, $Z^{*}$,
are in $e$. The contributions to $\epsilon$ are calculated in SI
units. For each IR active mode, the atomic motions are described.
mode. }
\label{table:BZ40PhonTable}
\end{table}

\begin{figure}
\includegraphics[width=3.0in]{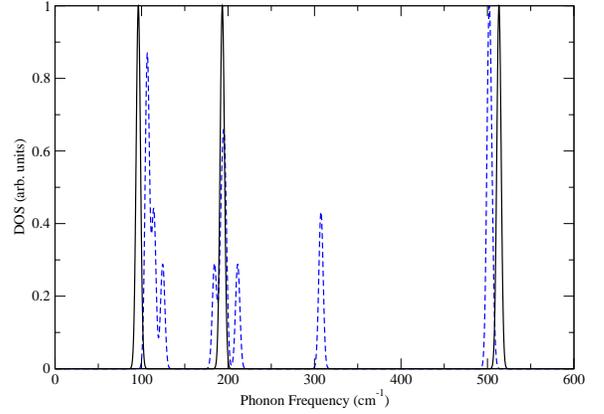}
\caption{{Gamma-point phonon DOS of the high-symmetry BaZrO$_{3}$
structure, solid black line, and the fully relaxed pseudo-cubic
supercell, dashed blue line.The most important changes are the upshift
in Ba-O mode frequencies (from 96 cm$^{-1}$ to 100-130 cm$^{-1}$) and the
appearance of new modes around 300 cm$^{-1}$.}}
\label{fig:BZDOS}
\end{figure}

Analysis of the low-symmetry (Table~\ref{table:BZ40PhonTable} and
Figure~\ref{fig:BZDOS}) phonon modes show that there are more
IR-active phonon modes than in the high-symmetry cell. The IR-active
modes are no longer of pure $x$, $y$ or $z$ character. However, they
can still be distinguished as Ba-O (Last mode), Zr-O (Slater mode) or
O$_6$ based.  Likewise, the frequency ranking of the modes is still
lowest for Ba-O, higher for Zr-O and highest for the O$_6$.  The Ba-O
modes are clustered around 115~cm$^{-1}$, about 10$\%$ higher in
frequency than for the high-symmetry cell.  The energies of the Zr-O
and the O$_{6}$ modes are essentially unchanged from their respective
5-atom cell values. Interestingly, a new band of modes is generated by
this structure around 300~cm$^{-1}$. do not contribute as much to
$\epsilon_{\mu}$ as the three previously mentioned sets of modes.

Comparison of the $\epsilon_\mu$ values in Tables I and II shows that
the contribution of the mainly Ba modes to $\epsilon$ has decreased by
half, whereas the contribution of the Zr modes and the O$_{6}$ modes
has remained the same. The new modes around 300 cm$^{-1}$ do not
contribute significantly. The total $\epsilon_{\rm \mu}$ is 45, in
contrast with 60 calculated for the 5-atom cell.  We find that
$\epsilon_{\infty}$ is still 4.94, and the total $\epsilon$ is 50, in
agreement with experimental data. Thus, we find that it is necessary
to include the O$_{6}$ tilt to correctly represent the dielectric
response of BZ at low temperature.

We now discuss the physical reasons behind the changes in the
dielectric response induced by octahedral tilting.  As noted above,
the weaker dielectric response of the low-symmetry BaZrO$_3$ structure
is caused by the 40\% drop in the contribution of the low frequency
Last mode.  To some extent, this effect can be rationalized by the
rattling cation model.  As the O$_6$ cages tilt, the O atoms are
brought closer to the Ba atom.  This reduces the volume of the
Ba-O$_{12}$ cage leaving less room for Ba to rattle around in response
to an applied electric field. More precisely, the shorter Ba-O
distances lead to a stiffening of the potential energy surface felt by
the Ba. This gives rise to a smaller dielectric response, as
$\epsilon_{\mu}$ is proportional to 1/$\omega^2$.  However,
examination of Ba-O modes shows that $\omega$ shifts up by about 10\%.
This should lead to about 27\% decrease in $\epsilon_\mu$, accounting
for about half of the Ba-O mode $\epsilon_\mu$ change.  The other
factor that weakens the dielectric response is the decrease in the
mode effective charge $Z^{*}_{\mu}$, which is significantly smaller
for the low-symmetry structure than for the 5-atom high-symmetry
structure. Since Eqns. 3 and 4 shows that $\epsilon_{\mu}$ is
proportional to the square of $Z^{*}_{\mu\alpha}$, distribution of
$Z^{*}_{\mu}$ over multiple modes leads to a 20\% decrease in
$\epsilon_\mu$.

Comparison of the Ba-O phonon eigenvectors for the two structures
shows that induced Zr off-centering is smaller for the Last mode in
the 40-atom low symmetry structure than in the 5-atom high-symmetry
mode.  It is this reduction in the induced Zr displacement that gives
rise to the smaller $Z^{*}_{\mu}$.  Physically, the smaller induced Zr
displacement points to a weakened coupling between Ba and Zr
displacements.  The coupling between Ba and Zr displacement is most
likely due to Pauli repulsion~\cite{Grinberg02p909}.  In the
low-symmetry structure O$_{6}$ tilting as well as increased O motion
in $x$ and $y$ directions in the Last mode eigenvector moves an O atom
between Ba and Zr. This partially screens the Ba-Zr interactions,
leading to a smaller $Z^{*}_{\mu}$.

\section{\label{sec:level1} Conclusions}
We have presented calculations that reveal interesting structural
and dielectric properties in the seemingly simple material
BaZrO$_{3}$. At zero temperature, the high-symmetry model proved
inadequate for calculating the dielectric constant.  Large supercells
were used in order to find the equilibrium structure, which exhibits
octahedral tilts. These tilts have a significant effect on the low
frequency Ba-O Last mode and its contribution to the dielectric
constant. Allowing octahdral tilts in the first principles modeling of
BaZrO$_{3}$ resolves the discrepancy between recent theoretical and
experimental investigations of low temperature dielectric
constant. Moreover, the understanding of how octahedral tilitng can
change dielectric response should give insight into alloying
strategies for future dielectric materials design.

\section{\label{sec:level1} Acknowledgments}
This work was supported by the Office of Naval Research under Grant
No. N-000014-00-1-0372, the Center for Piezoelectric Design, a GAAAN
fellowship from the University of Pennsylvania, and the IMI Program of
the National Science Foundation under Award No. DMR 0409848. We also
acknowledge the support of the DoD HPCMO, DURIP and NSF CRIF program,
Grant No. CHE-0131132.

\section{\label{sec:level1} References}
\bibliography{thebibliography}

\end{document}